\documentclass[twocolumn]{aastex61}
\usepackage{graphicx}
\usepackage{hyperref}
\usepackage{amssymb,amsmath}
\usepackage{url}
\usepackage{natbib}

\begin{document}
\title{Old but still warm: Far-UV detection of PSR B0950+08\footnote{Based on observations made with the NASA/ESA {\sl Hubble Space Telescope}, obtained at the Space Telescope Science Institute, which is operated by the Association of Universities for Research in Astronomy, Inc., under NASA contract NAS 5-26555. These observations are associated with program \#13783.}
}

\author{G.\ G.\ Pavlov}
\affiliation{Pennsylvania State University, Department of Astronomy \& Astrophysics, 525 Davey Lab., University Park, PA 16802; ggp1@psu.edu}
\author{B.\ Rangelov}
\affiliation{Texas State University,
Department of Physics,
601 University Drive, San Marcos, TX 78666}
\author{O.\ Kargaltsev}
\affiliation{George Washington University, Department of Physics, 725 21st St, NW, Washington, DC 20052}
\author{A.\ Reisenegger}
\affiliation{Instituto de Astrof\'{i}sica, Pontificia Universidad Cat\'{o}lica de Chile, Av.\ Vicu\~{n}a Mackenna 4860, Macul, Santiago, Chile}
\author{S.\ Guillot}
\affiliation{Instituto de Astrof\'{i}sica, Pontificia Universidad Cat\'{o}lica de Chile, Av.\ Vicu\~{n}a Mackenna 4860, Macul, Santiago, Chile}
\author{C.\ Reyes}
\affiliation{Instituto de Astrof\'{i}sica, Pontificia Universidad Cat\'{o}lica de Chile, Av.\ Vicu\~{n}a Mackenna 4860, Macul, Santiago, Chile}


\begin{abstract}
 We report on a {\sl Hubble Space Telescope} detection
of the 
nearby,
old pulsar B0950+08
($d\simeq 262$ pc, spin-down age 17.5 Myr)
 in two far-ultraviolet (FUV) bands.
We measured the mean flux densities $\bar{f}_\nu = 109\pm 6$ nJy
and $83\pm 14$ nJy in the F125LP and F140LP filters (pivot
wavelengths 1438 and 1528 \AA). 
Using the FUV data together with  previously obtained optical-UV
data, we conclude that the optical-FUV spectrum consists of two
components -- a nonthermal (presumably magnetospheric) power-law
spectrum ($f_\nu\propto \nu^\alpha$) with slope $\alpha\sim -1.2$
and a thermal spectrum emitted from the bulk of the neutron star 
surface with a temperature in the range of
(1--$3)\times 10^5$ K,
 depending on 
interstellar extinction and neutron star radius. 
These temperatures are much higher than predicted by neutron star
cooling models for such an old pulsar,
which means that some heating mechanisms 
operate in 
neutron stars. 
A plausible mechanism responsible
for the high temperature of PSR B0950+08 is 
the interaction of vortex lines of the faster rotating neutron
superfluid
 with the slower rotating normal matter in the inner 
neutron star crust (vortex creep heating).
\end{abstract}

\keywords{pulsars: individual (PSR\,B0950+08 = PSR\,J0953+0755) --- stars: neutron --- stars: ultraviolet}

\section{INTRODUCTION}
Born extremely hot, neutron stars (NSs) lose their thermal energy via neutrino and photon
emission. The cooling rate
 is determined by the state and composition of the NS interiors, which
are still poorly known. Therefore, the study of the thermal evolution
of NSs is 
an important tool for understanding the fundamental properties of
matter. In particular, the comparison of the surface temperatures,
$T_5\equiv T_\infty/10^5\,{\rm K}= 5$--20 ($T_\infty$ is the temperature as 
measured by a distant observer),
of young and middle-aged ($\tau\lesssim 1$ Myr)
NSs, measured in X-rays (e.g., \citealt{Pavlov2002}; \citealt{Deluca2005}) with model cooling curves $T_\infty(\tau)$ has constrained 
the properties of superfluidity
of the super-dense matter and perhaps even the masses of some isolated NSs
(\citealt{Yakovlev2004}; \citealt{Page2009}).

While the cooling of young NSs in the neutrino-dominated cooling era 
($\tau\lesssim 1$ Myr)
has been reasonably well investigated, the thermal evolution of older NSs remains virtually
unexplored. If a NS just cools passively, then its surface temperature is expected to drop
very fast in the photon-dominated cooling era ($\tau\gtrsim 1$ Myr), going below $10^4$ K at $\tau\sim 10$
Myr. However, it has long been recognized that various heating processes may slow (or even
reverse) the cooling (e.g., \citealt{Gonzalez2010}, and references therein).
The first observational evidence of heating of old NSs was obtained
by \citet{Kargaltsev2004} from {\sl Hubble Space Telescope} ({\sl HST}) observations of the 
 nearest millisecond (recycled)
pulsar J0437--4715, whose characteristic (spin-down) age\footnote{
Since 
the true 
ages of old pulsars are unknown, 
their spin-down ages,
$\tau_{\rm sd}\equiv P/(2\dot{P})$, are commonly used as 
age estimates.} 
$\tau_{\rm sd} 
\approx 7$ Gyr. 
These authors found that its far-UV (FUV) emission
is thermal, corresponding to a NS surface temperature 
$T_5 \sim 1$--2,
which was later confirmed by \citet{Durant2012}.
This high temperature was explained by
\citet{Fernandez2005} as caused by ``rotochemical heating'' due
to 
composition changes and accompanying non-equilibrium Urca reactions
(such as neutron beta decays) forced by the density
increase as the centrifugal force decreases in the course of NS spindown
\citep{Reisenegger1995}.

In addition to the rotochemical heating, other NS heating mechanisms
have been
proposed.
For instance, ``frictional heating''
due to interaction of vortex lines of the faster rotating neutron
superfluid with the slower rotating normal matter in the inner NS crust
should be most relevant for ``classical''
(non-recycled) pulsars with much longer periods and stronger magnetic
fields \citep{Alpar1984, Shibazaki1989, Larson1999, Gonzalez2010}.

To probe the thermal evolution of old NSs, including both
classical and recycled pulsars, we initiated an {\sl HST} program
\#\,17378.  First results from this program were reported by
\citet{Rangelov2017} who 
analyzed the observations of the solitary millisecond pulsar J2124--3358
and found that its FUV emission is likely thermal, corresponding
to a temperature $T_5\sim 0.5$--2.
Here we report the results of our observation of another target
of that program, the old classical pulsar B0950+08.

PSR B0950+08 (= J0953+0755; B0950 hereafter) is a solitary 
radio pulsar (no $\gamma$-ray emission detected)
 with the period $P=253$\,ms, spin-down energy loss rate
$\dot{E}=5.6\times 10^{32}$\,erg\,s$^{-1}$, 
spin-down age $\tau_{\rm sd}=
17.5$\,Myr, and surface magnetic field $B\sim 2.4\times 10^{11}$\,G 
\citep{Manchester2005}. B0950 has a parallax distance $d=262\pm5$\,pc
and an accurately measured proper motion,  $\mu_\alpha \cos\delta = -2.09 \pm 0.08$\,mas\,yr$^{-1}$, $\mu_\delta = 29.46\pm0.07$\,mas\,yr$^{-1}$,
corresponding to the transverse velocity 
$V_\perp=36.6\pm$0.7\,km\,s$^{-1}$ \citep{Brisken2002}.

X-ray emission from B0950 was barely detected with the {\sl Einstein} IPC \citep{Seward1988} and {\sl ROSAT} PSPC \citep{Manning1994}. 
Much deeper {\sl XMM-Newton} EPIC observations of B0950 were carried out 
in 2002.
\citet{Becker2004} described the phase-integrated
 X-ray spectrum with an absorbed power-law (PL)
 model 
($f_E\propto E^{1-\Gamma}$, where $f_E$ is the energy flux density)
  with  photon index
$\Gamma=1.92^{+0.14}_{-0.12}$, absorbing hydrogen column density $N_{H,20}\equiv N_H/(10^{20}\,{\rm cm}^{-2})=2.6^{+2.7}_{-2.4}$,
and unabsorbed flux $F_{\rm 0.5-10\,keV}^{\rm unabs}=
8.7^{+1.0}_{-0.9}\times 10^{-14}$\,erg\,cm$^{-2}$\,s$^{-1}$,
corresponding to the luminosity\footnote{This luminosity value is estimated
as $L=4\pi d^2 F^{\rm unabs}$, with $d=262$\,pc, as well as other luminosities
in this paper. The luminosity quoted by
\citet{Becker2004} is a factor of 4 lower.}
 $L_{\rm 0.5-10\,keV}=7.1^{+0.9}_{-0.7}\times 10^{29}$\,erg\,s$^{-1}$.
These authors 
also reported pulsations with one and two peaks per period in the 2--10
and 0.3--2\,keV bands, respectively.
\citet{Zavlin2004} suggested that the 
additional peak at lower energies can be explained by the contribution of thermal
polar cap emission.
They interpreted the same data as a combination of
nonthermal 
(presumably magnetospheric) 
emission with a PL spectrum ($\Gamma=1.3\pm0.1$, 
$L_{\rm 0.2-10\,keV}=
(1.0\pm 0.1)\times 10^{30}$\,erg\,s$^{-1}$) and 
thermal emission from heated polar caps covered with a hydrogen atmosphere,
which dominates at $E\lesssim 0.7$\,keV. Using hydrogen atmosphere models
\citep{Pavlov1995},
they estimated the effective temperature, radius, and
 bolometric luminosity of polar caps: 
$T_{\rm pc}\approx 1$ MK, $R_{\rm pc}\approx 250$\,m, 
$L_{\rm pc}\approx 3\times 10^{29}$\,erg\,s$^{-1}$,
at the hydrogen column density restricted to the range $2<N_{H,20}<4$.

Optical-UV emission from B0950 was discovered by \citet{Pavlov1996} in
{\sl HST} observations with the Faint Object Camera (FOC) in a long-pass
filter F130LP (pivot wavelength 3438\,\AA, ${\rm FWHM} \approx 2000$\,\AA).
They measured a mean spectral flux density $\bar{f}_\nu =51\pm 3$\,nJy in
this filter. 
\citet{Pavlov1996} considered the 
possibility that this could be, at least partly, thermal emission from
the entire NS surface, which gives an upper limit 
$T_5\lesssim 3$ on the
brightness temperature (for the apparent NS radius $R_\infty=13$\,km, the modern
distance estimate, $d=262$\,pc, and negligibly low interstellar extinction).

Observations 
with the 
Subaru telescope 
allowed \citet{Zharikov2002} to measure the flux density of $60\pm 9$\,nJy in 
the B filter.
Observations with the
VLT/FORS1 telescope 
\citep{Zharikov2004} gave
 the B, V, R and I 
flux densities of $60\pm 17$, $45\pm 6$, $91\pm8$ and $78\pm 11$
nJy, respectively. Despite the large scatter of the flux values
(likely caused by a contamination, particularly in the I and R bands,
 from a 
very red, possibly extended object at about $1''$ north of the pulsar;
see Figure 2 in \citealt{Zharikov2004}),
it is clear that the
optical emission from B0950 is nonthermal. 
Fitting these flux densities with a PL model, \citet{Zharikov2004} obtained 
a spectral slope $\alpha = -0.65\pm 0.40$
($f_\nu\propto \nu^{\alpha}$, $\alpha=1-\Gamma $). Figure 3 of that
paper suggests that the earlier detected emission in the F130LP filter
was mostly nonthermal, although some contribution from a thermal component
could not be excluded. \citet{Zavlin2004} have shown that the VLT/FORS1 
spectrum and the nonthermal component of the X-ray spectrum are consistent
with a single PL model, with $\Gamma\approx1.3$--1.4. 
For this model, the
broadband nonthermal luminosity is $L_{\rm 1\,eV - 10\,keV}^{\rm nonth}=1.1\times 10^{30}$\,erg\,s$^{-1} = 1.8\times 10^{-3} \dot{E}$, of which the optical
luminosity is a small fraction (e.g., $L_{\rm 1-5\,eV} = 5.2\times 10^{27}$\,erg\,s$^{-1} = 4.7\times 10^{-3} L_{\rm 1\,eV - 10\,keV}^{\rm nonth}$). 
Using this interpretation,
\citet{Zavlin2004} 
suggested an upper limit on the brightness temperature in the optical-UV: 
$T_5\lesssim 1.5$, for $R_\infty=13$\,km, $\alpha = -0.35$, and
reddening $E(B-V)=0.05$.

Thus, it still remains unclear how hot is the bulk of the B0950 NS surface.
To measure or constrain it, we observed this pulsar in the FUV
 with the {\sl HST}.
\section{Observations}
B0950 was observed with the Solar-Blind Channel (SBC) of the Advanced Camera for Surveys (ACS)
in two visits that occurred on 2016 January 29 (one orbit) and 2016 February 4 (two orbits). 
We split the observations in two 
visits
(see Table 1)
 to reduce the
instrumental background (``thermal glow'',  
caused by heating of the detector), which rises dramatically with the length of
time the detector is turned on\footnote{See \url{http://www.stsci.edu/hst/acs/documents/handbooks/current/c04_detector6.html\#325708}.}.
The SBC has a nominal field of view $\approx 34''\times 30''$. 
During each orbit the data were taken in
 long-pass filters F125LP and F140LP,
with  pivot
wavelengths 1438 and 1528 \AA, respectively
(see Figure 1).
 The F125LP images were taken in
the Earth shadow to minimize the geocoronal FUV background.
The F140LP filter, which cuts off bright geocoronal oxygen lines, was used in
non-shadow parts of the orbit.
\begin{figure}[h]
\begin{center}
\includegraphics[scale=0.61]{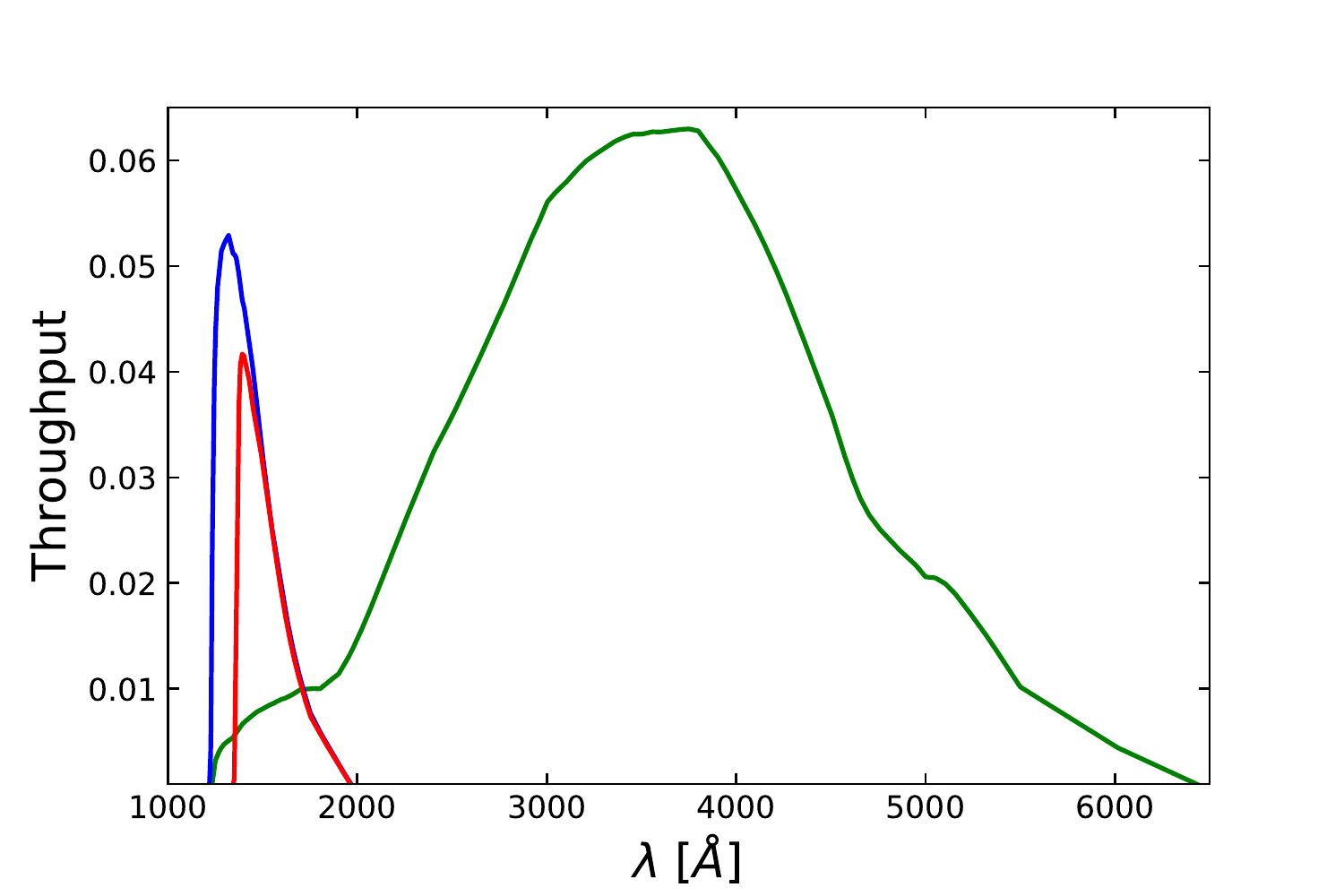}
\caption{Throughputs
of the {\sl HST} filters SBC/F125LP (blue), SBC/F140LP (red) and FOC/F130LP (green) used in the observations of B0950. \label{throughputs}
}
\end{center}
\end{figure}
\section{Data analysis and results}
For the data analysis, we used the pipeline-calibrated data files 
stored the Mikulski Archive for Space Telescopes (MAST\footnote{See \url{http://archive.stsci.edu/}}). The MAST images are coaligned and corrected for geometric distortion 
using the DrizzlePac {\tt AstroDrizzle} task\footnote{ 
\url{http://drizzlepac.stsci.edu/}}.
\subsection{FUV images}
Figure 2 
shows SBC/F125LP images of the B0950 field obtained in
the two visits.
 In each of the images we see a 
point-like source\footnote{
Zoomed-in images of the source obtained in separate visits
 show asymmetric extensions 
with lengths
$\lesssim 0\farcs2$
(see insets in Figure 2). Since the directions of these extensions are different
in different orbits, we conclude that they are due to instrumental effects.
}
 near the image
center and two extended objects, presumably galaxies.
 We also see that 
\begin{figure*}[ht!]
\begin{center}
\includegraphics[scale=0.52]{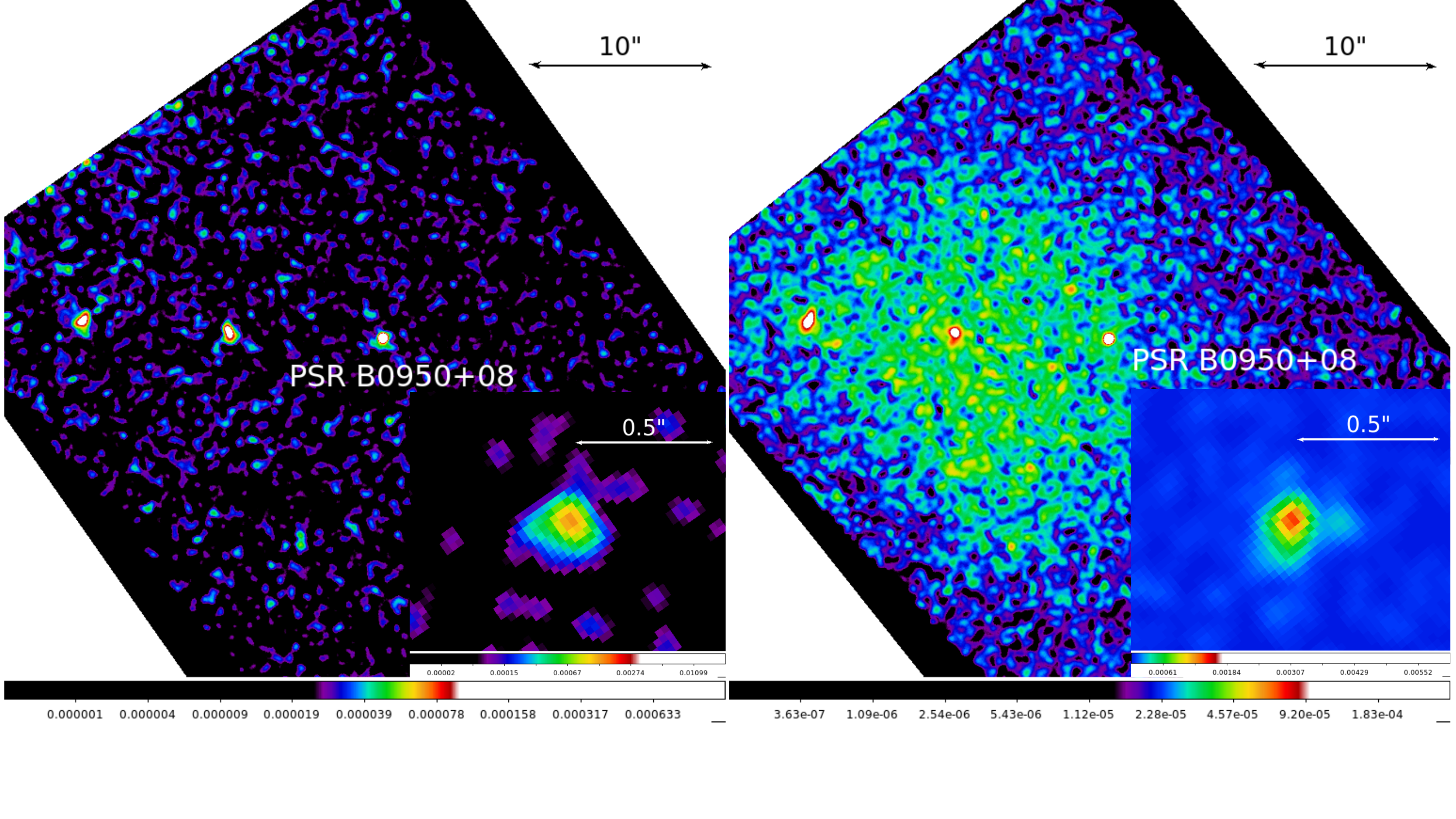}
\caption{SBC/F125LP images of B0950 and its surroundings for Visit 1
(one {\sl HST} orbit; left)
and Visit 2 (two orbits; right), smoothed with a 0\farcs1 width Gaussian.
North is up, East to the left.
The diffuse ``emission'' in the Visit 2 image is the instrumental ``thermal
glow'' caused by detector heating
during the second orbit of that visit.
The $\approx 1\farcs2 \times 1\farcs0$  insets show the
zoomed-in images of the pulsar and its immediate vicinity.
\label{fuv-images}
}
\end{center}
\end{figure*}
the SBC thermal glow background, 
indiscernible in the
one-orbit Visit 1, is 
much brighter in 
the two-orbit Visit 2 because the detector became hotter in the second orbit
of this visit.

The nominal coordinates of the point source
in the MAST images are $\alpha= 09$:53:09.347, 
$\delta=
07$:55:35.79 in Visit 1 and $\alpha= 09$:53:09.343, $\delta=
07$:55:35.74 in Visit 2 (centroiding uncertainty about a few times 0\farcs01).
The corresponding source positions differ by 1\farcs05 and 1\farcs06 
 from  the expected radio pulsar position,
$\alpha=09$:53:09.3054(19), $\delta=07$:55:36.64(8), at the time of observation
(MJD 57\,419). Such offsets, due to telescope pointing errors, are 
not unusual in {\sl HST} observations\footnote{See, e.g., \url{http://www.stsci.edu/hst/acs/documents/isrs/isr0506.pdf}}. In addition,
no 
field objects were seen in the pulsar vicinity in
the deep 
{\sl HST} FOC/F130LP  and Subaru B-filter 
observations.
Moreover, the optical-UV spectrum of the candidate pulsar counterpart
(see below) is quite unusual for a field star but
natural for a pulsar. It means that the detected
object is indeed an FUV counterpart of B0950.

We know from the VLT observations that there is a 
 background object
(likely a distant galaxy) 
very
close to the B0950 position, $\approx 0\farcs9$ from the pulsar in
2001 January \citep{Zharikov2004}. Its expected distance from the pulsar
became $\approx 0\farcs4$--0\farcs5 by the time of our observations (assuming
the object's proper motion is negligible).
However, this object is very faint and extremely red (not seen by Subaru and VLT in the B band).
Therefore, it should not make any contribution in FUV.
\subsection{Photometry}
For the pulsar photometry, we used 
``drizzled'' 
MAST images 
for each of the visits,
and the Python aperture photometry tool {\tt Photutils}. The pixel scale in 
the drizzled 
 images
is 0\farcs025.
To find the optimal source aperture, we calculated the signal-to-noise
ratio ($S/N$) as a function of radius of circular aperture centered on the
brightest pixel. For the F125LP images,
we chose the $r=10$ pixels 
source aperture
which provides $S/N\approx 10$ and 14 for Visit 1 and Visit 2, respectively.
We extracted 
the background from
an annulus with the inner and outer radii of 30 and 70 pixels (area $A_b=7.854$ arcsec$^2$). 
For the shallow
 F140LP images, the $S/N$ dependence on $r$ is rather noisy, but the same
$r=10$ pixels source aperture provides $S/N\approx 3.7$ and 4.6 (for Visits 1
and 2) close to their maximum values.

\begin{deluxetable*}{cccccccc}
\tablecaption{Photometry of PSR B0950+08}
\label{table:photometry}
\tablehead{
\colhead{Filter} & \colhead{Visit} & \colhead{$t_{\rm exp}$\tablenotemark{a}} &
\colhead{$N_{t}$\tablenotemark{b}} &
\colhead{$N_{ b}$\tablenotemark{c}} & \colhead{$N_{s}$\tablenotemark{d}} & \colhead{$C_{s}$\tablenotemark{e}} &
\colhead{$\bar{f}_\nu$\tablenotemark{f}} \\
\colhead{} & \colhead{} & \colhead{ks} &
\colhead{cnts} & \colhead{cnts} & \colhead{cnts} & \colhead{cnts\,ks$^{-1}$} & \colhead{nJy}
}
\startdata
F125LP & 1 & 1.836     &
114 & 477 &102 & $84.3\pm 8.8$ &
$100\pm11$ \\
F125LP & 2 &
3.692 &
268 & 1382 & 233 & $95.8\pm 6.7$ &
$114\pm 8$ \\
F125LP & 1+2 & 5.528      &
382 & 1859 & 336& $92.0\pm 5.4$ & $109.4\pm 6.4$ \\
F140LP & 1 & 0.592       &
19 & 110 & 16.2 & $41.5\pm
11.1$ & $88\pm 24$ \\
F140LP & 2&  1.182      &
43 & 532 & 29.8 & $38.2\pm
8.3$ & $81\pm18$
 \\
F140LP & 1+2& 1.774  &
62 & 642 & 46.0& $39.0\pm 6.7$ & $83\pm 14$ \\
\hline
\enddata
\tablenotetext{a}{Exposure time.}
\tablenotetext{b}{Total number of counts in the
$A_s=0.196$ arcsec$^2$ source aperture.}
\tablenotetext{c}{Background counts in the $A_b=7.854$ arcsec$^2$ annulus.}
\tablenotetext{d}{Net source counts in the $A_s=0.196$ arcsec$^2$ aperture.}
\tablenotetext{e}{Aperture-corrected source count rate.}
\tablenotetext{f}{Mean flux density of the source (see text).}
\end{deluxetable*}

Using the numbers of counts, $N_t$ and $N_b$, in the source and background
apertures (see Table 1),
we calculated the number of source counts, $N_s=N_t - (A_s/A_b)N_b$,
and its uncertainty, $\delta N_s = [N_t + (A_s/A_b)^2 N_b]^{1/2}$, for
each of the filters and visits. We corrected 
the corresponding count rates
for the finite
aperture size using the encircled energy fractions of 0.66 and 0.665
for the F125LP and F140LP filters, respectively\footnote{The encircled energy fractions
were calculated using Table 2 in the Instrument Science Report ACS 2016-05 by Avila \& Chiaberge, \url{http://www.stsci.edu/hst/acs/documents/isrs/isr1605.pdf}. The aperture size is sufficiently large to neglect the effect of the 
artificial ``extensions'' of the pulsar image (see footnote $^6$ and Figure 2)
on the encircled energy fraction. }.

We used the aperture-corrected count rates, $C_s$ in Table 1,
to calculate the mean flux densities,
\begin{equation}
\bar{f}_\nu \equiv \left[\int f_\nu T(\nu) \nu^{-1} d\nu\right] \left[\int T(\nu) \nu^{-1} d\nu\right]^{-1} = {\cal P}_\nu C_s,
\label{meanfluxdensity}
\end{equation}
where $f_\nu$ is the energy flux density,
 $T(\nu)$ is the filter throughput as a function of frequency,
and ${\cal P}_\nu$ is the conversion factor\footnote{Note that ${\cal P}_\nu$
(and hence the $\bar{f}_\nu$ value for
a given $C_s$) 
do not depend on the shape of $f_\nu$. In a similar manner, one can define $\bar{f}_\lambda =
{\cal P}_\lambda C_s$, where ${\cal P}_\lambda = {\cal P}_\nu c \lambda_{\rm piv}^{-2}$ (`photflam' header keyword in calibrated {\sl HST} data), $\lambda_{\rm piv}^2=\left[\int T(\lambda) \lambda\,d\lambda\right] \left[\int T(\lambda) \lambda^{-1} d\lambda\right]^{-1}$ is the pivot wavelength squared.}
   (${\cal P}_\nu =1.19$
and 
2.12 nJy\,ks\,cnt$^{-1}$ for the F125LP and F140LP filters, respectively).

\begin{figure}[t]
\begin{center}
\includegraphics[scale=0.62]{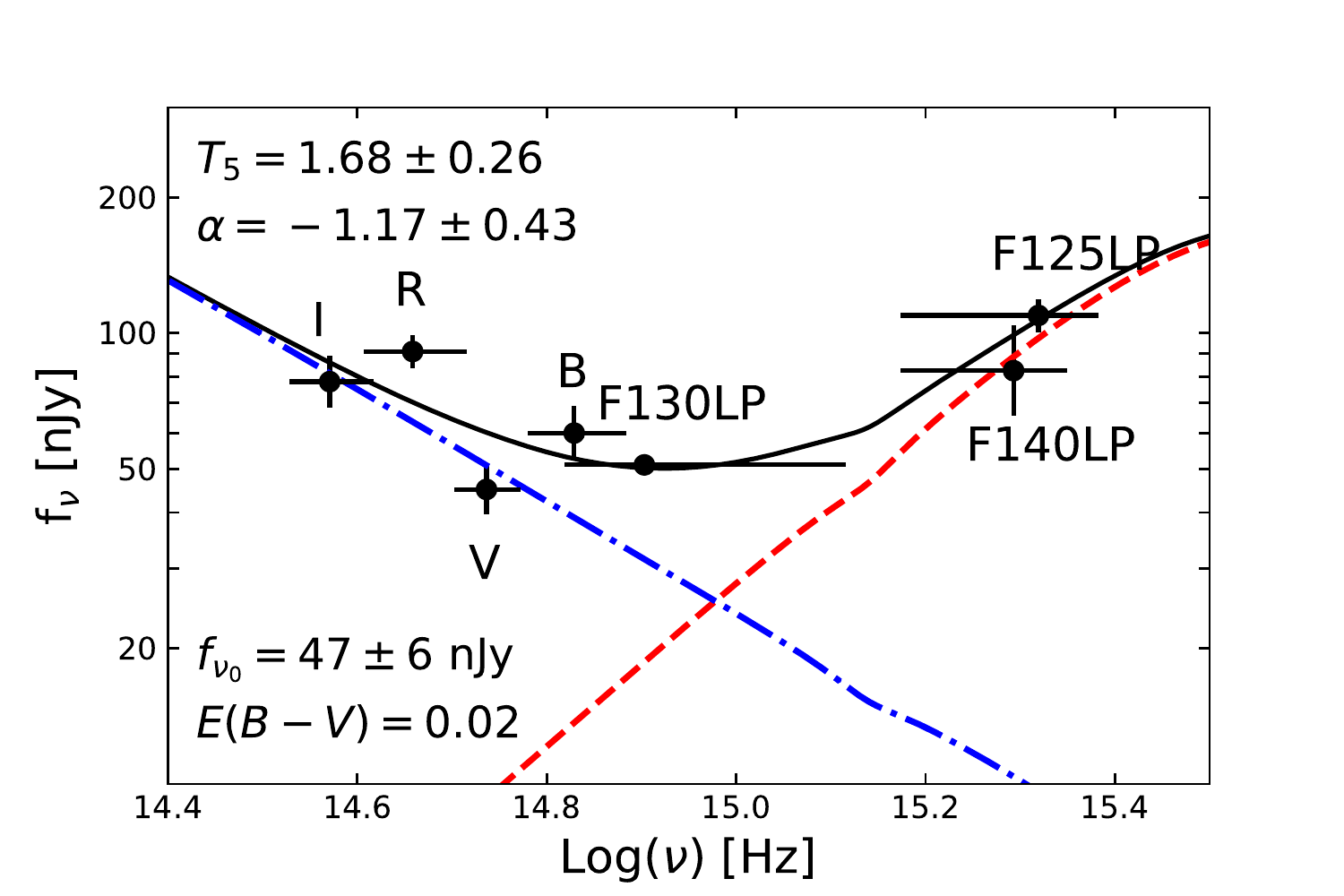}
\caption{Spectral fit for the {\sl HST} (F125LP, F140LP, F130LP), Subaru (B)
and VLT (V, R, I) data
with the PL+BB model for $E(B-V)=0.02$, $R_{12}/d_{262}=1$.
The horizontal bars show the filter widths at half maxima of their throughputs,
the mean flux values are plotted at the pivot frequencies of the filters.
\label{spectral-fit}
}
\end{center}
\end{figure}

\subsection{Spectral fits}
We used our photometry results and the previous {\sl HST}, Subaru and VLT observations
\citep{Pavlov1996, Zharikov2002, Zharikov2004} to fit model spectra to the data.
Contrary to the VLT optical spectrum,
the overall broadband spectrum in the 1250--9000 \AA\ range (see, e.g., 
Figure 3)
cannot be described by a single PL model because of the relatively high
FUV flux (e.g, at $E(B-V)=0.02$ the best-fit PL model, $\alpha=+0.4\pm0.3$, corresponds 
to an unacceptably large reduced $\chi_\nu^2=10.6$ for 5 degrees of freedom [dof]). 
The FUV excess can be naturally interpreted as thermal emission from the
NS surface.
Therefore, we fit the data with a two-component PL+blackbody (BB) model (see, e.g.,
\citealt{Pavlov1997}; \citealt{Kargaltsev2007}):
\begin{equation}
f_\nu = \left[f_{\nu_0}\left(\frac{\nu}{\nu_0}\right)^\alpha + 
\frac{R_\infty^2}{d^2}\pi B_\nu(T_\infty)\right]\times 10^{-0.4 A_\nu},
\label{model}
\end{equation}
where 
$\nu_0$ is the reference frequency (we chose $\nu_0= 6\times 10^{14}$\,Hz),
$d=262 d_{262}$\,pc is the distance, $B_\nu(T_\infty)=(2h\nu^3/c^2) [\exp(h\nu/kT_\infty) -1]^{-1}$ is the Planck function, the temperature $T_\infty=10^5T_5$\,K and the radius $R_\infty=12 R_{12}$\,km are as measured by a distant observer.

The extinction coefficient $A_\nu$ is proportional to the color excess $E(B-V)$
(dust reddening), which can be crudely
 estimated using the hydrogen column density: 
$E(B-V)= (0.0146\pm 0.0006) N_{H,20}$\,mag (e.g., \citealt{Guver2009}). 
The $N_H$ values obtained from
X-ray spectral fits, $N_{H,20} \sim 1$--4, are rather uncertain; they 
correspond to $E(B-V)\sim 0.014$--0.06. 
The correlation between $N_H$ and
pulsar dispersion measure, $N_{H,20} = 0.30^{+0.13}_{-0.09}\,{\rm DM}$ \citep{He2013}, gives $N_{H,20}\approx 0.6$--1.3, $E(B-V)\approx 0.01$--0.02 (${\rm DM}=2.97$\,pc\,cm$^{-3}$ for
B0950). The color excess can also be estimated from 3D reddening maps\footnote{See \url{http://stilism.obspm.fr/} and \url{http://argonaut.skymaps.info/}}:
$E(B-V)=0.015\pm0.008$ and $E(B-V)=
0.004$--0.030 in the maps described by \citet{Lallement2014} and \citet{Green2015}, respectively.
Thus, a plausible range of the color excess is $E(B-V)=0.01$--0.03, but
higher values, up to 0.06, are not firmly excluded.
We will explore the conservative range,
$E(B-V)=0.01$--0.06 
in our fits, using the extinction curve `Milky Way' from
Table 3 by \citet{Clayton2003} to 
calculate the extinction $A_\nu$. 

The model given by Equation (\ref{model}) has 5 parameters:
$f_{\nu_0}$, $\alpha$, $R_\infty/d$, $T_\infty$, and $E(B-V)$. Because of the small number
of data points and a correlation between the parameters, 
we have to fix some of
them to obtain constrained fits. We chose to fit $f_{\nu_0}$, $\alpha$, and $T_\infty$  
at fixed 
values of $E(B-V)$ and $R_\infty/d=1.48\times 10^{-15} R_{12}/d_{262}$.
An example of such a fit at plausible values of $E(B-V)=0.02$ and 
$R_{12}/d_{262}=1$ is shown in Figure 3.
We see from Figure 3
that the PL component gives the main contribution
to the optical flux, while the FUV flux is determined by the BB component
with $T_5=1.68\pm 0.26$ and bolometric luminosity $L_{\rm bol}^\infty
= 8.2^{+6.4}_{-4.0} \times 10^{29}$ erg\,s$^{-1}$.
The quality of this fit (and of the fits at other $E(B-V)$ and $R_\infty/d$ values) is
poor (reduced $\chi_\nu^2 = 3.8$ at 4 dof)
because of the large scatter of the optical 
(Subaru and VLT) fluxes, likely caused by large
systematic errors in 
correcting for the contribution of the nearby
red object (see Figure 2 in \citealt{Zharikov2004} and Section 1),
unaccounted in the 
flux uncertainties.
The PL normalization, $f_{\nu_0} = 47\pm 6$\,nJy,
and particularly the slope,
 $\alpha=-1.17\pm 0.43$, are rather uncertain, due to the same reason.
The slope is steeper than that obtained by \citet{Zharikov2004},
 $\alpha=-0.65\pm 0.40$, because those authors 
did not include the
 thermal component, 
which gives a considerable contribution to the
F130LP flux in our fits.

\begin{figure}[ht!]
\begin{center}
\includegraphics[scale=0.62]{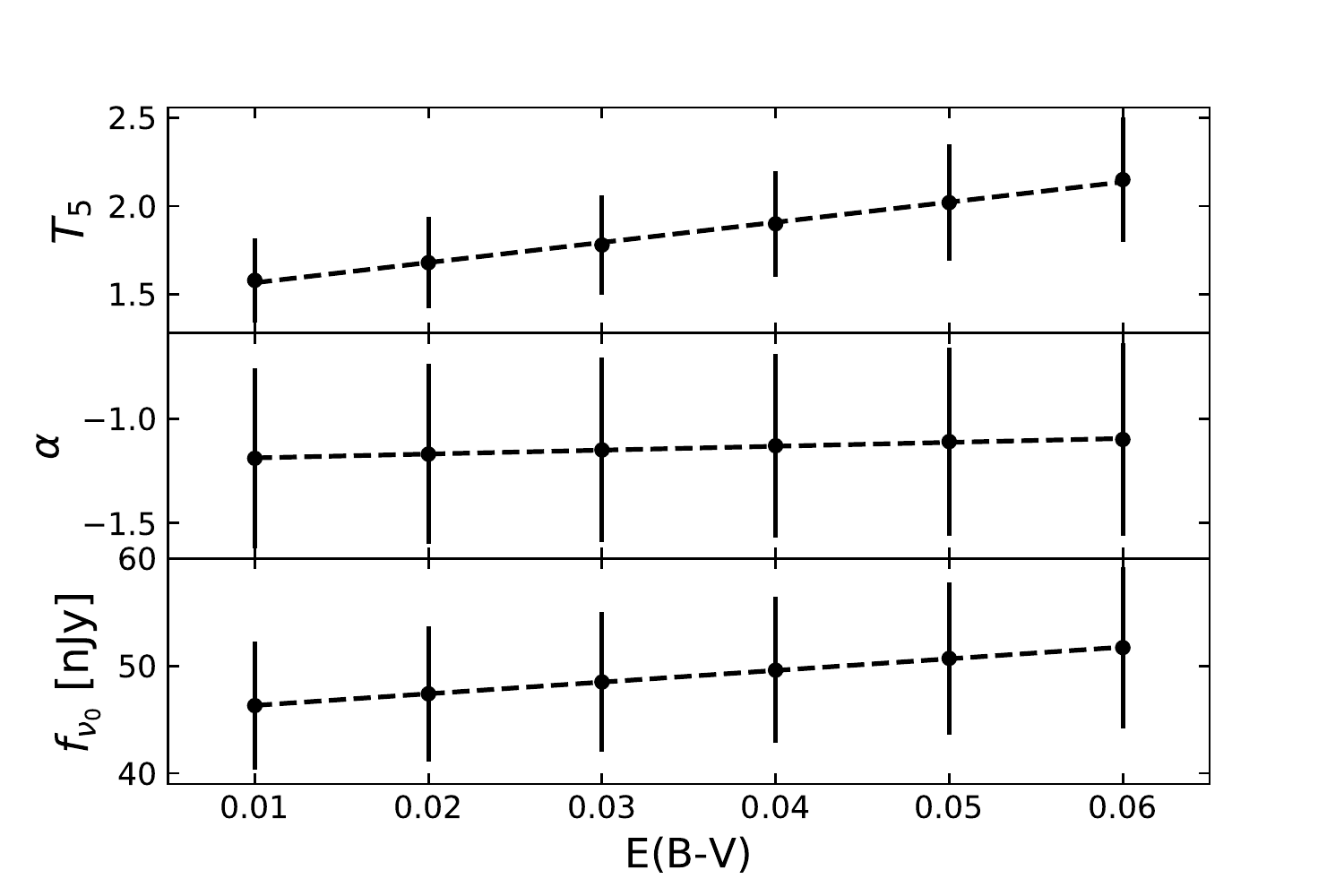}
\caption{Dependences of the fitting parameters on color excess
and their linear approximations (see Equations (3)--(5)), for 
$R_{12}/d_{263}=1$.
\label{reddening-dependence}
}
\end{center}
\end{figure}

The dependences of $T_5$, $\alpha$ and $f_{\nu_0}$ on the reddening 
$E(B-V)$ for 
$R_{12}/d_{262}=1$ are shown in Figure~4.
They can be approximated by linear functions: 
\begin{eqnarray}
T_5 &= &1.45\pm 0.22 + (11.4\pm 2.16)\,E(B-V), \\
\alpha & = & -1.21\pm 0.42 +(1.8\pm 0.6)\, E(B-V),\\
f_{\nu_0}&=&[45.2\pm5.7 +(108\pm29)\,E(B-V)]\,\,{\rm nJy}.
\end{eqnarray}
We see from Figure 4
that, for the chosen $R_{12}/d_{262}=1$ and $E(B-V)=0.01$--0.06,
 a conservative range for the BB temperature
is $1.3 \lesssim T_5 \lesssim 2.5$
($1.3\lesssim T_5\lesssim 2.1$ for a more plausible extinction range
$E(B-V) = 0.01$--0.03).

\begin{figure}[t]
\begin{center}
\includegraphics[scale=0.62]{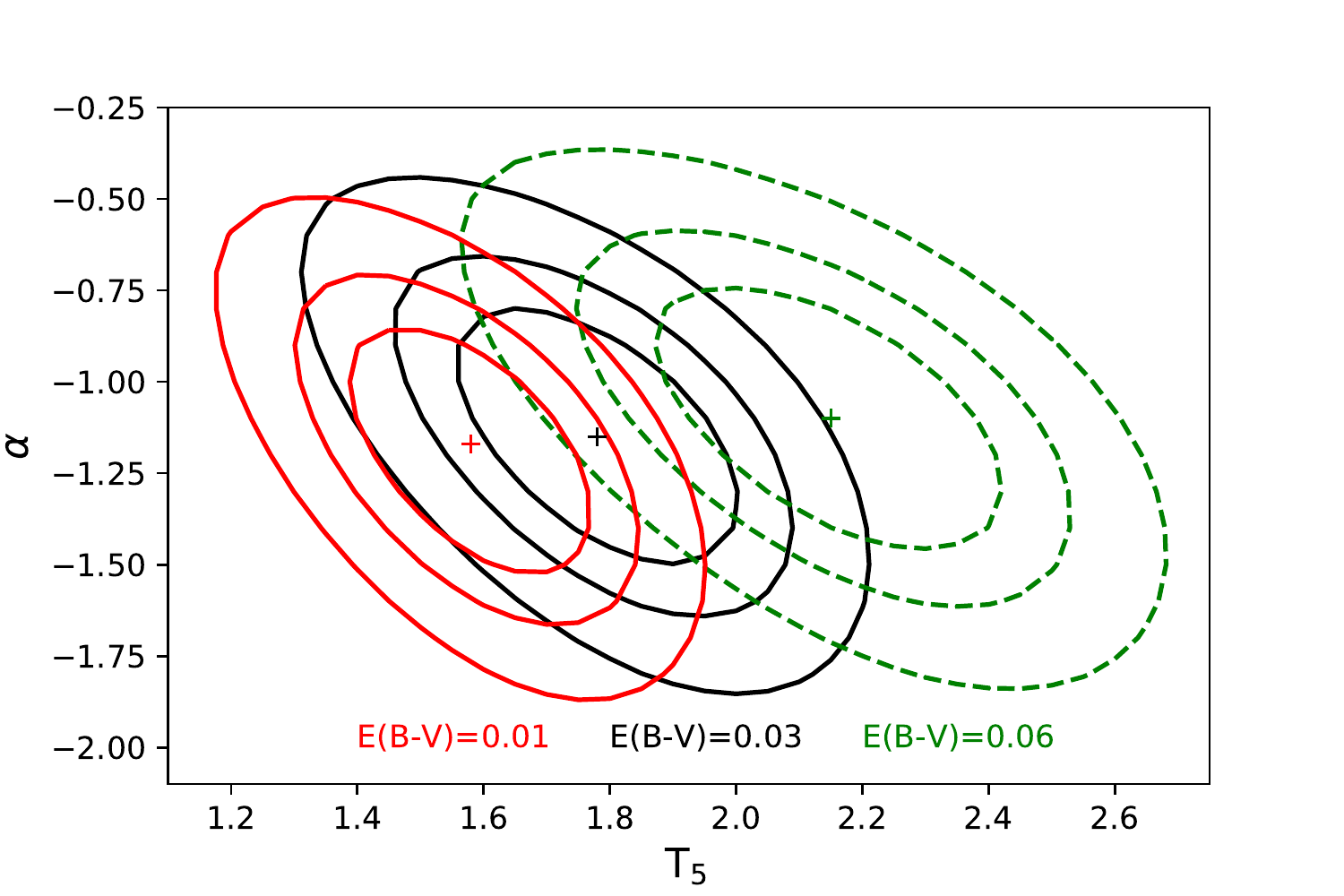}
\caption{Confidence contours 
in the $T_\infty$-$\alpha$
plane for $R_{12}/d_{262}=1$ and three values of $E(B-V)$.
The contours are lines of constant values of 
$\chi^2_{\rm min}+\Delta \chi^2$, where $\chi^2_{\rm min}$ is the minimum
value of $\chi^2$ at a given $E(B-V)$, and 
$\Delta\chi^2=2.3$, 4.6, and 9.2 for the 68\%, 90\%, and 99\% 
confidence levels, respectively, for two parameters of interest. 
The third parameter,
$f_{\nu_0}$, was 
 varied to minimize $\chi^2$ at each point of the $T_\infty$-$\alpha$ grid.
  The solid 
red and black contours
correspond to the minimum 
and maximum 
values of the
most plausible range of the color excess, $E(B-V)=0.01$ and 0.03, respectively;
the dashed green countours correspond to the conservative upper limit,
$E(B-V)=0.06$.
\label{T-alpha_contours}
}
\end{center}
\end{figure}

An additional uncertainty of the fitting parameters is caused by
the uncertainty of the $R_\infty/d$ ratio (mostly due to uncertain NS radius).
Similar fits for a few $R_{12}/d_{262}$ values from the plausible range
$0.9< R_{12}/d_{262} < 1.2$ show that the best-fit temperature
and its upper and lower bounds are 
proportional to\footnote{This dependence is weaker than $T\propto (R_\infty/d)^{-2}$, expected for the case when all the other parameters are fixed, 
because $\alpha$ becomes more negative with increasing $R_\infty/d$,
which results in higher fitting temperatures (see Figure 5).}
 $(R_{12}/d_{262})^{-1.5}$,
and the most 
conservative range of temperatures is
$1.0 \lesssim T_5\lesssim 3.5$ ($1.0 \lesssim T_5\lesssim 3.0$ for the
more plausible range of color excess).
The temperature range is rather broad, but we stress that because of the
 $T$-$\alpha$ anti-correlation 
 and the large uncertainty of
$\alpha$ in the current fits
(see Figures 4 and 5), much more stringent constraints on $T$ could
be obtained after more accurate measuremnts of NIR-optical fluxes, which
would require observations with high angular resolution.

Additional constraints on the temperature upper limit could be obtained from
the X-ray data, but 
this limit depends on the interpretation of the X-ray emission, 
which is still controversial. For instance, assuming that the 0.3--10 keV
spectrum is an abosrbed PL, 
we found $3 \sigma$ upper limits $T_5 <3.1$ and $T_5<3.3$ at
$N_{H,20}=2$ and $N_{H,20}=4$, respectively (for $R_\infty = 13$ km).
These limits are close to the upper bound on the temperature range obtained from the FUV-optical data.
Assuming that the X-ray spectrum includes  both the magnetospheric (PL)
component and the thermal component emitted by polar caps covered with
hydrogen atmosphere, \citet{Zavlin2004} obtained a lower value
of the upper limit, $T_5 < 1.5$,
for $R_\infty = 13$ km, $E(B-V)=0.05$. However, this 
model-dependent estimate was obtained
assuming a constant slope of the optical through X-ray spectrum,
 $\alpha = -0.35$, much more gradual than $\alpha= -1.12\pm0.45$ obtained in
our FUV-optical fits at the same
$E(B-V)$. In addition, the polar cap contribution and
hence that upper limit depend on the poorly known 
orientation of the pulsar's spin axis and the magnetic axis inclination
as well as on the mass-to-radius ratio. These quantities could be more
accurately estimated from phase-resolved spectral analysis of 
data obtained from a deeper X-ray observation, which would also be
useful to establish the nature of the X-ray emission.

\subsection{Upper limit on extended emission in pulsar vicinity}
Although old and not very powerful, B0950 should emit a pulsar wind and be accompanied by a pulsar wind nebula (PWN). Its radiation could be emitted by 
shocked relativistic wind and/or shocked interstellar medium (ISM). 
Although FUV is not a wavelength range in which PWNe are usually looked for,
the recent detections of FUV bow shocks from the millisecond pulsars 
J0437--4715 \citep{Rangelov2016} and J2124--3358 \citep{Rangelov2017} suggest that 
FUV PWNe could be detected from other nearby pulsars.
Therefore, we looked for extended emission in the vicinity of B0950.

A characteristic PWN size can be estimated by balancing the PW pressure 
with the pressure exerted by the ambient ISM,
including the ram pressure 
caused by the pulsar motion
(see, e.g., Equation (1.1) in \citealt{Kargaltsev2017}). In the case of
B0950 the characteristic size can vary from a few tenths of arc second to a few
arc seconds, depending on the local ISM properties and the pulsar
velocity.
Inspection of our FUV images did not show extended emission around
the pulsar.
We measured the upper limit on its surface brightness (specific intensity)
 by  sampling  the  background  counts  from  13  different circular regions with $r=1''$ within $10''$ from the pulsar,  calculating  the  standard  deviation from  the  mean  value and converting it to the mean flux upper limit.   
In the F125LP low-background image of Visit 1
 we found a standard deviation of 21.9 counts ks$^{-1}$ for the count rate in the $r=1''$ circles. It translates into
a $3\sigma$ upper limit of 8.3\,nJy\,arcsec$^{-2}$ for the specific intensity.

\section{Discussion}
Our FUV observations of B0950, 
analyzed together with the previous optical-UV obervations,
have  shown that its optical radiation is nonthermal (presumably
emitted from the pulsar magnetosphere) while the FUV radiation is predominantly
thermal, most likely emitted from the bulk of the NS surface.
The discovery of the thermal emission from the 
$\sim$20 Myr 
old pulsar
with the brightness temperature
substantially higher than predicted by the NS cooling theories
is the main finding
of this work. 

The 
estimated temperature of B0950, $T_5= 1$--3,
is similar to those found for the millisecond pulsars J0437--4715 (\citealt{Kargaltsev2004}; \citealt{Durant2012}; \citealt{Guillot2016})
and J2124--3358 \citep{Rangelov2017},
hereafter J0437 and J2124. However, B0950 is a classical 
(non-recycled) pulsar, with a much longer 
rotation period, a stronger magnetic field, and a younger spin-down age, though 
substantially older than middle-aged ($\tau \sim 0.1$--1 Myr) NSs
whose temperatures are generally
compatible with 
the predictions of standard cooling mechanisms.

\citet{Gonzalez2010} analyzed various possible 
heating mechanisms for old 
NSs, finding that vortex creep and rotochemical heating
by non-equilibrium Urca reactions in the NS core were the only ones that could plausibly explain the fairly high temperature of J0437\footnote{
A variant of rotochemical heating through pycnonuclear reactions in the NS crust was later shown to be similarly effective for millisecond pulsars \citep{Gusakov2015}.}.
 The lower panels of Figures 4 and 5 in \citet{Gonzalez2010} show the expected thermal evolution of classical pulsars with a magnetic field $B=2.5\times 10^{11}$ G (nearly identical to that of B0950) for these two heating mechanisms, and the observational upper limits for several pulsars.
From those plots, we infer that the 
estimated temperature range for
B0950 is roughly consistent with their curves for vortex creep heating with either modified or direct Urca cooling, where the ``excess angular momentum'' parameter $J= 5.5\times 10^{43}$ erg s
 was chosen as the smallest compatible with the temperature measurement of J0437, which is also consistent with J2124. On the other hand, 
to explain the temperature of B0950 with rotochemical heating, we would
need to invoke pure modified Urca reactions and an implausibly short initial rotation period ($P_0\lesssim 10$ ms), in order to allow the pulsar to accumulate a large enough chemical imbalance by its current age. 
UV observations of other old classical pulsars, supplemented by deep optical and X-ray observations, would allow one to firmly establish the heating mechanism and take it into account in the models of thermal evolution of NSs. 
 
Regarding the PL optical spectrum, we note that 
if its slope
is as steep as found from our fits, $\alpha\approx -1.2\pm 0.5$, then the
extrapolation of the optical PL 
to higher energies underpredicts the X-ray flux, 
requiring a ``double break'' in the optical through X-ray spectrum.
Such behavior is similar to that observed in the middle-aged
pulsar B1055--52 (see Figure 3 in \citealt{Mignani2010}), but it is unusual for pulsars 
detected in both the X-ray and optical ranges, for which the extapolated
optical spectrum either overpredicts or matches the X-ray spectrum
\citep{Kargaltsev2007}.
The steep optical spectrum of B0950
 might suggest that different populations of relativistic particles are
responsible for the optical and X-ray nonthermal emission, but the 
estimated slope of the optical emission suffers from 
large systematic uncertainties.
To measure the slope of the optical component more accurately,
the pulsar should be
resolved from the nearby extended optical source, presumably a distant galaxy.
High-resolution observations in a few NIR-optical filters would not only determine the slope of the pulsar's optical
spectrum but would also allow one to
measure the NS surface temperature more accurately than it was possible
with the data available.

The detection of thermal emission from B0950, J0437, and J2124,
as well as the discovery of FUV bow shocks around J0437 and J2124,
demonstrate the great potential of UV observations of nearby pulsars.
Earlier UV-optical
observations of another type of NSs,
the so-called Thermally Emitting Isolated NSs, 
whose emission is
powered by the heat stored in the NS interiors rather than by the loss of
their rotational energy,
also showed a number of unexpected,
not fully understood properties \citep{Kaplan2011, Ertan2017}.
It would be very important to employ the unique
UV capabilities of the {\sl HST} for the study of a larger sample of isolated
NSs of various types.     

\begin{acknowledgements}
Support for program \#13783 was provided by NASA through a grant from the Space Telescope Science Institute, which is operated by the Association of Universities for Research in Astronomy, Inc., under NASA contract NAS 5-26555.
The work of AR and CR is supported by FONDECYT Regular Project 1150411, and that of SG by FONDECYT Postdoctoral Project 3150428.
We are grateful to Roberto Avila for very useful consultations on the ACS SBC detector
properties. We also thank Marten van Kerkwijk and Denis Gonz\`{a}lez-Caniulef
for valuable help in preparation of the observational proposal.
\end{acknowledgements}

\facility{{\sl HST}(ACS)}
\software{DrizzlePac (\url{http://drizzlepac.stsci.edu/}),
 Photutils}
\newpage
\bibliography{psr}
\end{document}